\documentclass[referee,a4paper,12pt,traditabstract]{swsc} 

\usepackage{graphicx}
\usepackage{txfonts}
\usepackage{subfigure}
\usepackage{epstopdf}
\usepackage[displaymath,mathlines]{lineno}
\usepackage[authoryear,round]{natbib}
\usepackage[backref]{hyperref}
\usepackage{url}

\hypersetup{colorlinks=true,citecolor=cyan,urlcolor=cyan,linkcolor=blue}

\begin{document} 


\title{On a limitation of Zeeman polarimetry and imperfect instrumentation in representing 
solar magnetic fields with weaker polarization signal}

\titlerunning{Zeeman polarimetry of magnetic fields with
weak polarization}
\authorrunning{A.A. Pevtsov et al.}

\author{Alexei A. Pevtsov\inst{1}
\and
Yang Liu\inst{2}
\and 
Ilpo Virtanen\inst{3}
\and 
Luca Bertello\inst{1}
\and
Kalevi Mursula\inst{3}
\and
K.D. Leka\inst{4, 5}
\and
Anna L.H. Hughes\inst{1}
}

\institute{National Solar Observatory, 
3665 Discovery Drive, 3rd Floor, 
Boulder, CO 80303 USA\\
 \email{\href{mailto:apevtsov@nso.edu}{apevtsov@nso.edu}}
 \and 
Stanford University, Stanford, CA, USA
\and
ReSoLVE Centre of Excellence, Space Climate research unit, University of Oulu, POB 3000, FIN-90014, Oulu, Finland
\and
NorthWest Research Associates, 3380 Mitchell Lane, Boulder, CO 80301 USA
\and
Institute for Space-Earth Environmental Research, Nagoya University, Furo-cho Chikusa-ku Nagoya, Aichi 464-8601 Japan
}


\abstract
{Full disk vector magnetic fields are used widely 
for developing better understanding of large-scale structure, morphology, and patterns of the solar magnetic field. The data are also important for modeling 
various solar phenomena.
However, observations of vector magnetic fields have one important limitation
that may affect the determination of the true magnetic field orientation. 
This limitation stems from our ability to interpret the differing character of the Zeeman polarization signals which arise from the photospheric line-of-sight vs. the
transverse components of the solar vector magnetic field, and is likely exacerbated by unresolved structure (non-unity fill fraction) as well as the disambiguation of the 180$^\circ$ degeneracy in the transverse-field azimuth.
Here we provide a description of this phenomenon, and discuss issues, which require additional investigation.}

\maketitle
%

\section{Introduction} \label{sec:intro}
Observations of magnetic fields provide key information for developing our 
understanding of the Sun's short-term (space weather) and long-term (space climate) activity and in predicting these effects on Earth. Synoptic full disk 
longitudinal magnetograms have existed since the late 1960s, and these data continue to 
serve as the primary input for space weather and space climate research and operational forecasts. By their nature, longitudinal magnetograms do not 
contain sufficient information to derive the true orientations of the magnetic-field 
vectors, and thus, require additional assumptions for physical interpretation. For example, ``radial 
field'' synoptic maps, which are widely used in space weather forecasting are created under 
the assumption that the true field is radial. 
Observations of vector Stokes polarimetery in principle 
 have the information necessary to fully reconstruct photospheric vector-magnetic-field maps, 
and efforts are underway to employ such data in operational space weather forecasts.

The earliest observations of vector magnetic fields in solar active regions were conducted at the Crimean 
Astrophysical Observatory in the early 1960s \citep{Stepanov.Severny1962,Severny1965}. By the early 1980s, a number of vector 
magnetographs were developed 
around the world, with the most prolific instruments operating in  Czechoslovakia, East Germany 
\citep{Pflug.Grigoriev1986}, Japan \citep[NAOJ,][]{Ichimoto.etal1993}, the Soviet Union 
(Crimean, Pulkovo and Sayan observatories), and the USA (NASA's Marshall Space Flight Center/MSFC, Mees Solar Observatory of University of Hawaii, High Altitude Observatory/HAO) \citep[for 
review, see individual articles in][]{Hagyard1985,Cacciani.etal1990}. 
All of these instruments had 
a limited field of view, typically about the size of an average active region.
Full disk vector 
magnetograms have been routinely observed since late 2003 by
the Vector Stokes Magnetograph (VSM) on the Synoptic Optical Long-term Investigation of the Sun (SOLIS) platform \citep{Keller.etal2003}.
Beginning in 2010, full disk vector magnetograms are available from the Helioseismic and Magnetic Imager 
\citep[HMI,][]{Scherrer.etal2012} on board the Solar Dynamics Observatory (SDO).

\begin{figure}
\includegraphics[width=1.0\columnwidth,clip=]{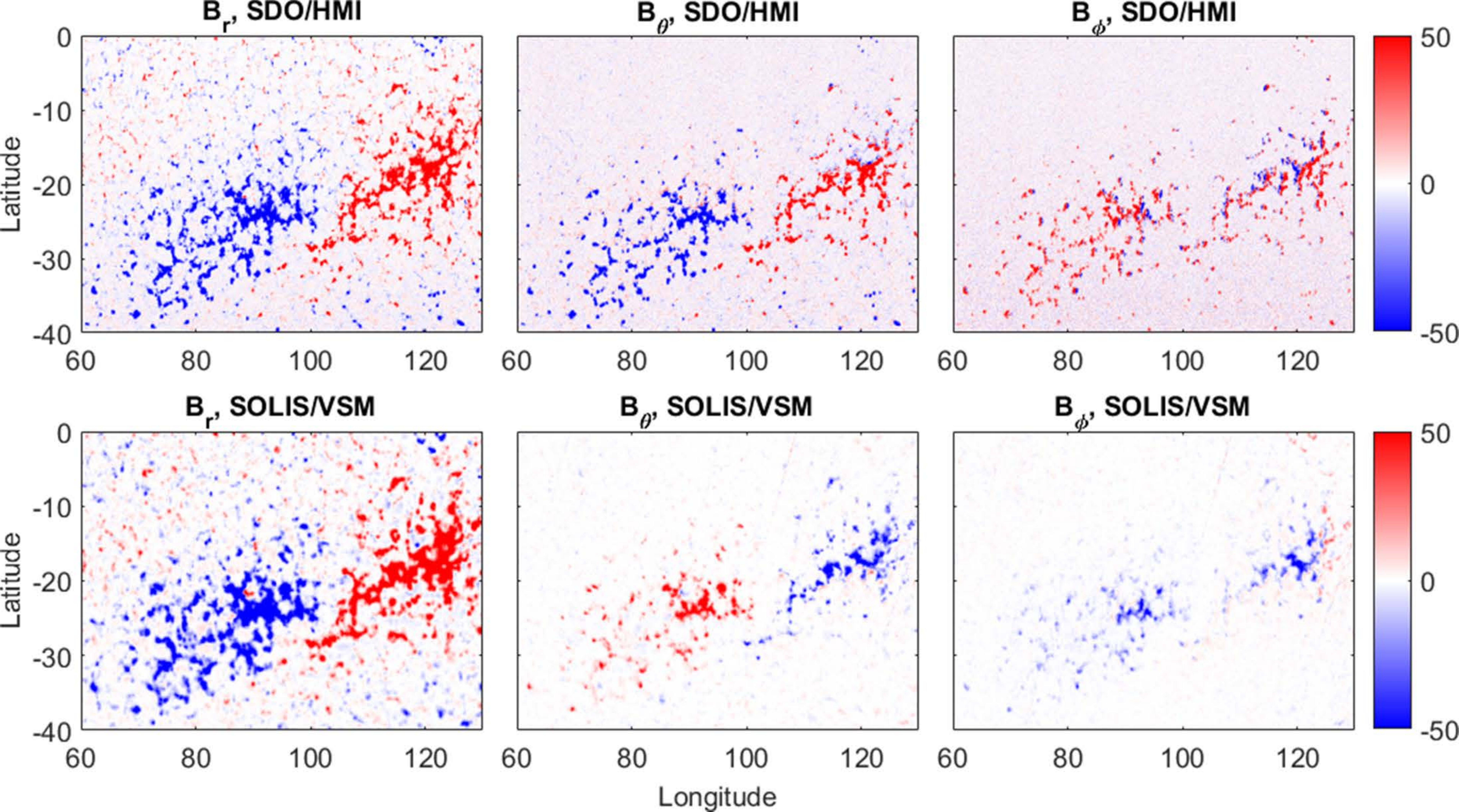}
\caption{Example of disagreement in orientation of B$_\theta$ and B$_\varphi$
Magnetic field vector components of an active region in the southern hemisphere, which passed the central meridian on 19 November 2015. Upper panels show radial (B$_r$), meridional (B$_\theta$) and zonal (B$_\varphi$) field from SDO/HMI and lower panels from SOLIS/VSM. Meridional (middle) and zonal (right) components depict opposite orientation. Red/blue represent positive/negative polarities scaled between
$\pm$ 50~G (see color bars on the right side of figure).
\label{fig:hmi_vsm}}
\end{figure}

On 23--26 January 2017, a working meeting on the ``Use of Vector 
Synoptic Maps for Modeling'' was held in Oulu, Finland with two follow up meetings  at the National Solar Observatory, Boulder, Colorado, USA (7--10 November 2017), and at the Max Planck Institute for Solar 
System Research,  G{\"o}ttingen, Germany (18-21 September, 2018).
At the first meeting, a direct comparison of vector magnetic field observations from SDO/HMI and SOLIS/VSM revealed disagreements in the orientation of meridional 
(B$_\theta$, North-South) and/or zonal (B$_\varphi$, East-West) components of the vector field in plage areas.
One example of this disagreement is shown in Figure \ref{fig:hmi_vsm}, which shows the three components of the magnetic field in VSM and HMI synoptic maps between heliographic latitudes -40$^\circ$ and 0$^\circ$ and Carrington longitudes 60$^\circ$ and 130$^\circ$.

This decaying active region (plage) was at the central meridian on 19 November 2015.
A synoptic map is used as an example because the averaging used in producing a synoptic map results in lower background noise, but very similar patterns are also seen in less-averaged full-disk data. 
Recent comparisons also indicate that the average large-scale zonal field B$_\varphi$ observed by SOLIS/VSM and derived from line-of-sight observations tends to disagree outside active regions \citep{Virtanen.etal2019}.
The investigation of this disagreement uncovered 
what we think may be an important limitation to
Zeeman vector polarimetry.
Here we present the results of our investigation into the discrepancies uncovered by that collaborative effort. In Section \ref{sec:data}, we introduce observations from two vector magnetographs and compare the vector field derived from these instruments and their standard data reduction algorithms. 
Sections \ref{sec:hypothesis}-\ref{sec:alpha} introduce our explanation of the observed disagreement, and in Section \ref{sec:discuss} we discuss the results of our findings.

\section{Vector magnetograms from different instruments and a comparative 
analysis} \label{sec:data}

Here, we employ full disk vector magnetograms from two instruments: the Vector Stokes Magnetograph (VSM) on the Synoptic Optical Long-term Investigations 
of the Sun (SOLIS) platform \citep{Keller.etal2003,Balasubramaniam.Pevtsov2011}, and the Helioseismic 
and Magnetic Imager \citep[HMI,][]{Scherrer.etal2012,Hoeksema.etal2014} on board the Solar Dynamics Observatory \citep[SDO,][]{Pesnell.etal2012}.

The HMI instrument is a filtergraph covering the full solar disk with $4096 \times 4096$ pixels.
The spatial resolution is about 1" with a 0.5" pixel size. The width of the
filter profiles is 7.6 pm. The spectral line used is Fe {\sc i} 617.3 nm,
which forms in the photosphere \citep{Norton.etal2006}. 
The Stokes parameters $(I, Q, U, V)$ are computed
from those measurements \citep{Couvidat.etal2016}, and are further inverted to retrieve the vector
magnetic field using a Milne-Eddington (ME) based inversion algorithm,
the Very Fast Inversion of the Stokes Vector \citep[VFISV][]{Borrero.etal2011,Centeno.etal2014}.
To suppress {p}-mode oscillations and to increase the signal-to-noise
ratio,  registered filtergrams are averaged over a certain time
before computing the Stokes vector. Currently a weighted average is computed every
720 seconds using data obtained over 1350 seconds by default; other averaging windows are available.

Inversion of the vector field has an unavoidable 180$^\circ$ ambiguity in
the azimuthal field direction.  Assumptions about the field must be made to
resolve the ambiguity. For all pixels in active regions, as well as for strong-field
pixels (where the S/N $>$3 in the transverse signal plus a 50~G buffer) in quiet Sun regions, the
azimuth is determined using a minimum energy algorithm \citep{Metcalf1994,Metcalf.etal2006,Leka.etal2009,Hoeksema.etal2014}.
The minimum-energy-method computation is time consuming for pixels where the signal
is dominated by
noise, so for weaker polarization regions, the $180^\circ$
ambiguity is solved using three quicker methods: a randomizing 
method (the option to add 180$^\circ$ is determined randomly), an acute-angle comparison 
to a potential field, and a method that provides the most 
radially-directed solution. More details can be found in \citet{Hoeksema.etal2014}. 
In this study, we used the random disambiguation for pixels with weaker linear 
polarization.

The VSM is a  spectrograph-based instrument, which observes full line profiles of the Fe {\sc i} 630.15 and 630.25~nm spectral lines, with a spectral sampling of 2.4~pm and pixel size of 1.0~$\times$~1.0 (1.14~$\times$~1.14 before
January 2010) arcseconds over a 2048~$\times$~2048 pixels field of view.
 To construct a full-disk magnetogram, 
the image of the Sun is scanned in the direction perpendicular to the spectrograph slit. At each scanning step,
the spectra for each pixel along the slit are recorded simultaneously. Each scanning step takes 
0.6~s, and a full disk magnetogram can be completed in about 20 minutes. The spectrograph slit is curved, 
which introduces geometric distortions to the image of the Sun. These distortions are corrected by 
shifting the position of each pixel in the final image to the closest integer position of the true pixel  
location in a round-Sun image. The maximum uncertainty in the position of a pixel does not exceed 
half-a-pixel, which is significantly smaller than typical atmospheric seeing for this groundbased 
instrument. The above correction procedure avoids ill-posed interpolation of full disk 
magnetograms, and it preserves the mapping of spectral information for each image pixel.

Similar to HMI, the observed profiles of Stokes {\rm Q}, 
{\rm U}, {\rm V}, and {\rm I} parameters are inverted using the VFISV code under the assumption of a standard Milne-Eddington stellar atmosphere. However, unlike HMI, VSM inversion includes 
the magnetic-field filling factor ($\alpha$) as an additional fit parameter, which represents the fraction of each instrument pixel filled by magnetized plasma.
For additional details about 
SOLIS/VSM inversion methods and pipeline, see \citet{Harker2017}. The 180$^\circ$ azimuthal ambiguity in the transverse field is resolved using the Very Fast 
Disambiguation Method \citep[VFDM,][]{Rudenko.Anfinogentov2014}. The VFDM has an 
accuracy almost as good as 
that of the minimum energy method (used for HMI disambiguation), but is much faster.
For a synoptic instrument such as VSM and HMI, the disambiguation is done automatically
as part of the pipeline data reduction. The pipeline reductions are optimized for 
``a good answer most of the time, in time for the next dataset'', and in some cases
may not return the best possible solution.

VSM data are from data processing level PROVER0 = 15.0511, 
which uses only one (Fe {\sc i} 630.15 nm) spectral line for the inversion.
In the following discussion we adopt the righthanded coordinate system (B$_r$,B$_\theta$,B$_\varphi$), which has been 
used in previous publications
\citep[e.g., ][]{Virtanen.etal2019}. Here the 
radial component (B$_r$) is positive when pointing  away  from  the  Sun,  the  meridional  component  (B$_\theta$) is  positive  southward, and  the  zonal  component  (B$_\varphi$)  is  positive westward.

\section{Effects of noise in the transverse fields on the derived vector-field orientation} \label{sec:hypothesis}

Let us now consider how properties of noise could affect the 
derived interpretation of the 
vector magnetic field. In all modern instruments, the derivation of vector 
magnetic fields is based on observations of full Stokes profiles in a 
selected spectral line 
sensitive to the effects of the magnetic field at the location of the formation of 
this line. Stokes {\rm I} represents the total intensity of light. 
Stokes {\rm V} is circular 
polarization (counter-/clock-wise), and Stokes {\rm Q} and {\rm U} 
represent two linear polarizations. The observed 
Stokes profiles are fitted by a model/synthetic line profiles in a 
process called ``inversion'', and the 
properties of the magnetic field (and other parameters, such as 
Doppler velocity, 
temperature, magnetic filling factor etc) are determined based on 
properties of the fitted line profiles.

The observed profiles of all Stokes parameters are affected by noise and other 
instrumental limitations. 
The photometric noise in Stokes {\rm Q} and {\rm U} is similar to that of
Stokes {\rm V}. However, the longitudinal (B$_{||}$) field is related 
linearly to circular polarization, while the 
relation between transverse field and linear polarization is quadratic (e.g., see Equation 18 in \citealt{Stenflo2013} and Equations 12c and 12d in \citealt{JeffriesMickey1991}), which results in a lower signal/noise in the latter for the same underlying field strength.
For example, in the weak-field approximation  
\citep{Ronan.etal1987,JLS89,JeffriesMickey1991}, 
\begin{equation}
B_{||} \approx \frac{C_{||}}{ \alpha}~{\rm V}
\label{eq1}
\end{equation}
\begin{equation}
B_\perp \approx \frac{C_\perp}{\sqrt\alpha} \sqrt[4]{{\rm Q}^2 + {\rm U}^2}.
\label{eq2}
\end{equation}

Based on radiative transfer 
modeling of the Fe {\sc i} 630.15 nm and assuming a filling factor of unity, $\alpha = 1$ (the entire pixel is filled with a field), \citet{Stenflo2013} estimated the coefficients 
of proportionality as $C_{||} \approx 29.4$ and $C_\perp \approx 184$.
(The coefficients are model dependent, but 
$C_\perp >> C_{||}$ independent of a model).

As a result, the noise measured in horizontal fields is typically larger 
than the amplitude of noise in the longitudinal field by a factor of 10-25. Moreover, unlike noise in 
B$_{||}$, which is distributed around zero, (B$_\perp \pm$~noise) is always positive 
\citep[e.g. see Appendix A in][]{Leka.Skumanich1999}.
For some specific magnetic configurations, this dichotomy may
systematically distort the true 
inclination of the vector magnetic field.

\begin{figure}
\hbox{\includegraphics[width=0.5\columnwidth,clip=]{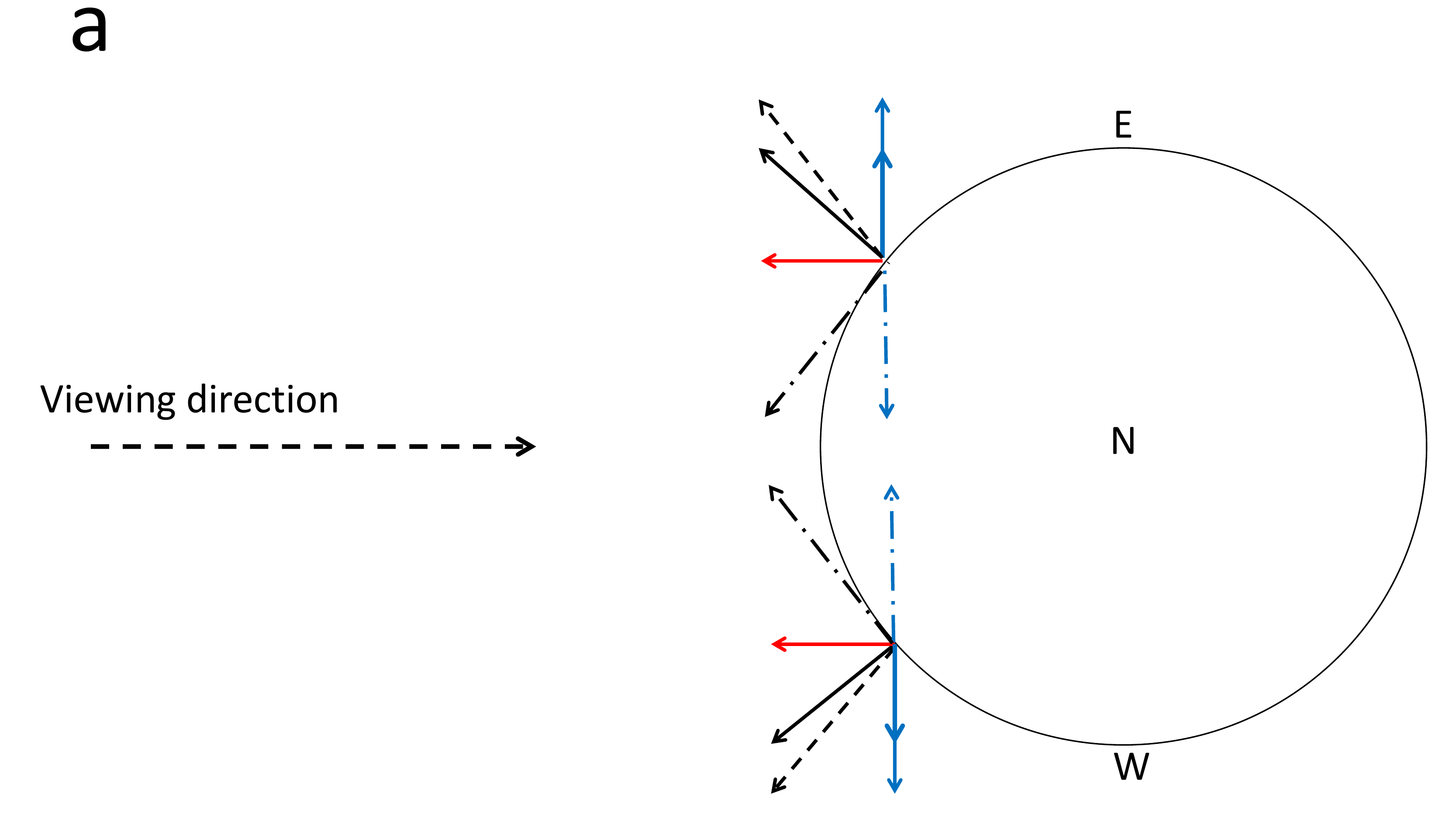}
\includegraphics[width=0.5\columnwidth,clip=]{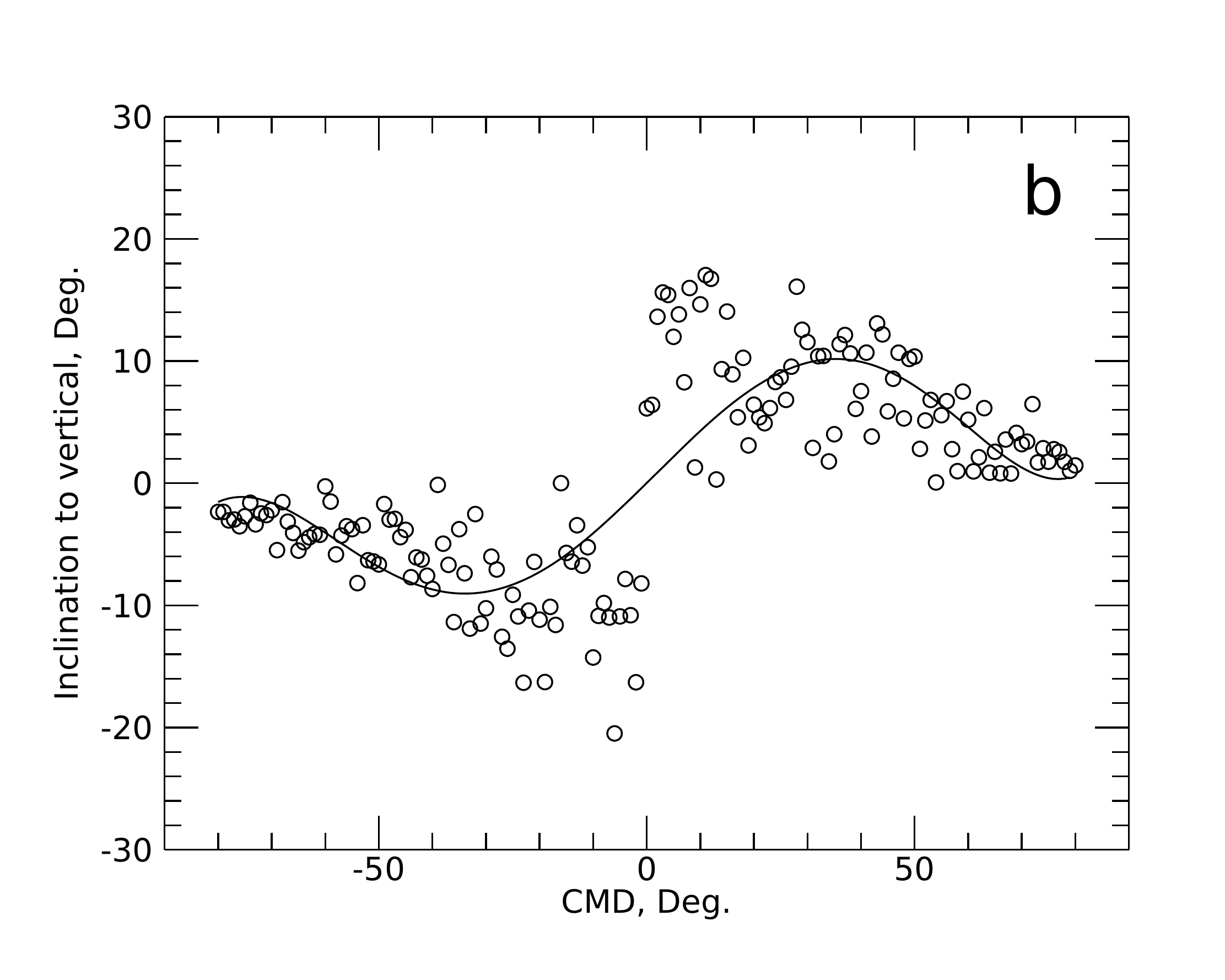}}
\caption{Effect of noise in the transverse field on the derived 
orientation of the vector field. 
(a) Black arrows represent the true vector 
field on the Sun (radial in this example). Red arrows represent the 
line-of-sight component and thick blue arrows are transverse 
components of the true vector field. Adding positive noise to 
the transverse field (thin blue arrows) makes the observed field 
(dashed arrow) systematically inclined in the direction away from 
the central meridian. Because of the 180$^\circ$ azimuthal ambiguity in the 
transverse field, the same transverse field may satisfy 
an alternative orientation (dashed-dotted arrow). The resulting 
orientation of the field vector is shown by a dashed arrow for both locations. Letters E, W, N mark approximate 
positions of solar East limb, West limb, and North as seen from 
Earth. 
(b) Difference between the true radial direction and vector 
orientation in the presence of noise, when azimuth disambiguation selects the most radial of two possible solutions (open circles). Solid line is a 6th degree polynomial fit.
\label{fig:demo}}
\end{figure}

Figure \ref{fig:demo}a  shows a theoretical example, where the true magnetic field is expected to be radial (black arrow). Observed at disc center, a radial  magnetic field is then determined by the observed line-of-sight component plus symmetric random noise in the azimuth direction, which is negligible on average. 
However, outside the central meridian as depicted in Figure \ref{fig:demo}a, a purely radial field contributes to both the line-of-sight component (red arrow) and the transverse component (blue arrow). 

In the presence of noise, the orientation of the magnetic vector will be determined by the observed (B$_{||}\pm \delta B_{||}$) and (B$_\perp \pm \delta B_\perp$) components, where $\delta B$ represents the amplitude of noise. Note that due to a quadratic contribution of Stokes \rm{Q} and \rm{U} to B$_\perp$ (see, Equation \ref{eq2}), (B$_\perp\pm\delta B_\perp$) is always positive (by definition, transverse field is always positive, with or without noise).

For pixels situated East of the central meridian (upper part of Figure \ref{fig:demo}a), the projection of the same (radial) vector 
would create a systematic non-zero transverse component, which in the presence of 
noise will only increase (due to B$_\perp \pm \delta B_\perp >$ 0).
Due to the 180-degree ambiguity in the azimuth of the transverse field, the vector 
field would have two possible orientations, and if the more radial (black solid arrow) option
is chosen, then the selected orientation of the magnetic field will have a systematic tilt in 
the direction away from the solar disk center. Figure \ref{fig:demo}b shows a difference
between the true radial field and one with $\delta B_\perp$ about 20~G added to the true transverse 
field. For this modeling exercise, we assumed that the true field strength of the vector field was 200~G, and $\delta B_{||}$ = 0 (no error in the longitudinal field), and $\delta B_\perp$ was
randomly generated within the range of about 0--100~G, with a mean about 50~G and a standard deviation of 20~G.
For pixels located closer 
to the disc center, we see an increase in scatter in the vector inclination relative to the true
radial direction. Most importantly, there is a systematic inclination of the vector field 
in the direction away from the central meridian. For a field strength of 200 G and 20~G noise
in transverse field, the inclination error could be up to 20 degrees (Figure \ref{fig:demo}b).

\begin{figure}
\includegraphics[width=1.0\columnwidth,clip=]{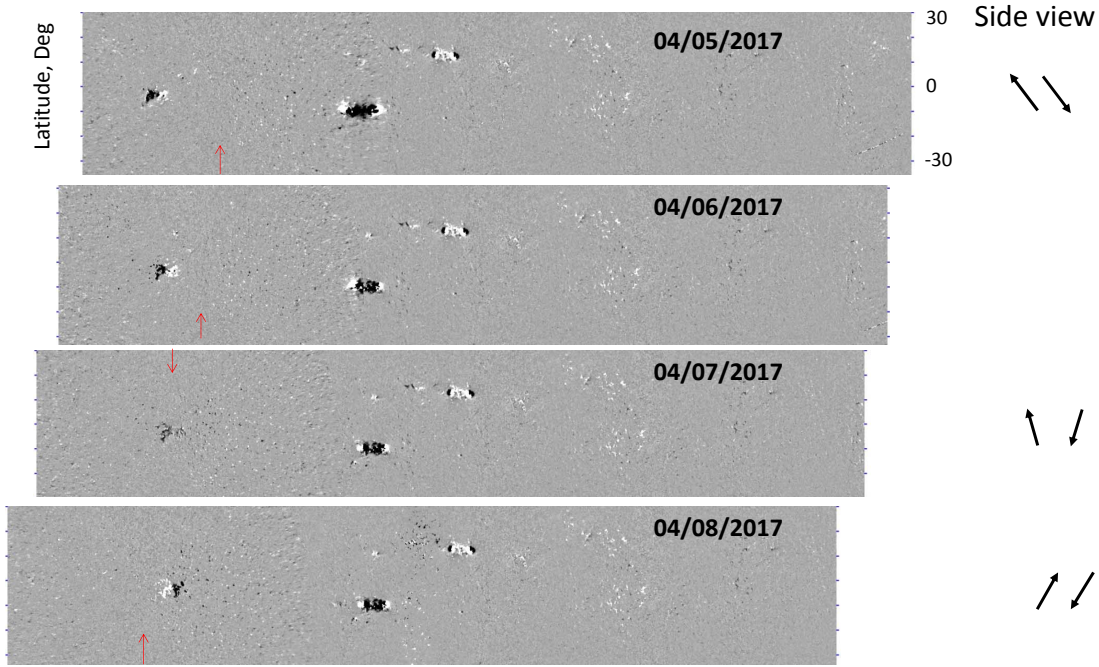}
\caption{(Left) Changes in the pattern of the zonal (East-West) vector-field component of a small bipolar region over its disk passage. Four panels show the so-called Near Real Time (NRT) synoptic maps. White/black halftones correspond to magnetic field directed towards the West/East. Each NRT map
covers 360 degrees in longitude (horizontal direction) and approximately $\pm$ 30 degrees in latitude (vertical direction). The most recent data are added onto the left side of synoptic maps. The dates of the most recent observations added to synoptic map are shown in the upper-right corner of each panel. For visual clarity, the synoptic charts are shifted to have the active regions aligned in the vertical direction. Small red arrows 
plotted in the left part of each panel correspond to the approximate location of the central meridian for the day of observations. Panels on the right show a schematic inclination pattern of the magnetic field vectors in a vertical East-West oriented plane (as if we were looking at an active region from the side). 
\label{fig:vsm_nrt}}
\end{figure}

Figure \ref{fig:vsm_nrt} demonstrates this effect in the disk-passage of a small bipolar region with negative leading and positive following polarity fluxes. 
It is clear that when this small active region is located to the East of central meridian, the horizontal 
magnetic field connecting the two polarities is directed westward in the leading (negative) polarity and 
eastward in the following (positive) polarity flux element. Only when the region is near the
central meridian does it become clear that the magnetic field in both polarity fluxes is close to radial,
with a very slight inclination in the direction away from each polarity. When the region is located 
West of central meridian, the pattern of the zonal field reverses as compared with its location East of central 
meridian. This behaviour is in perfect qualitative agreement with the explanation given in Figure \ref{fig:demo}.

\begin{figure}
\includegraphics[width=1.0\columnwidth,clip=]{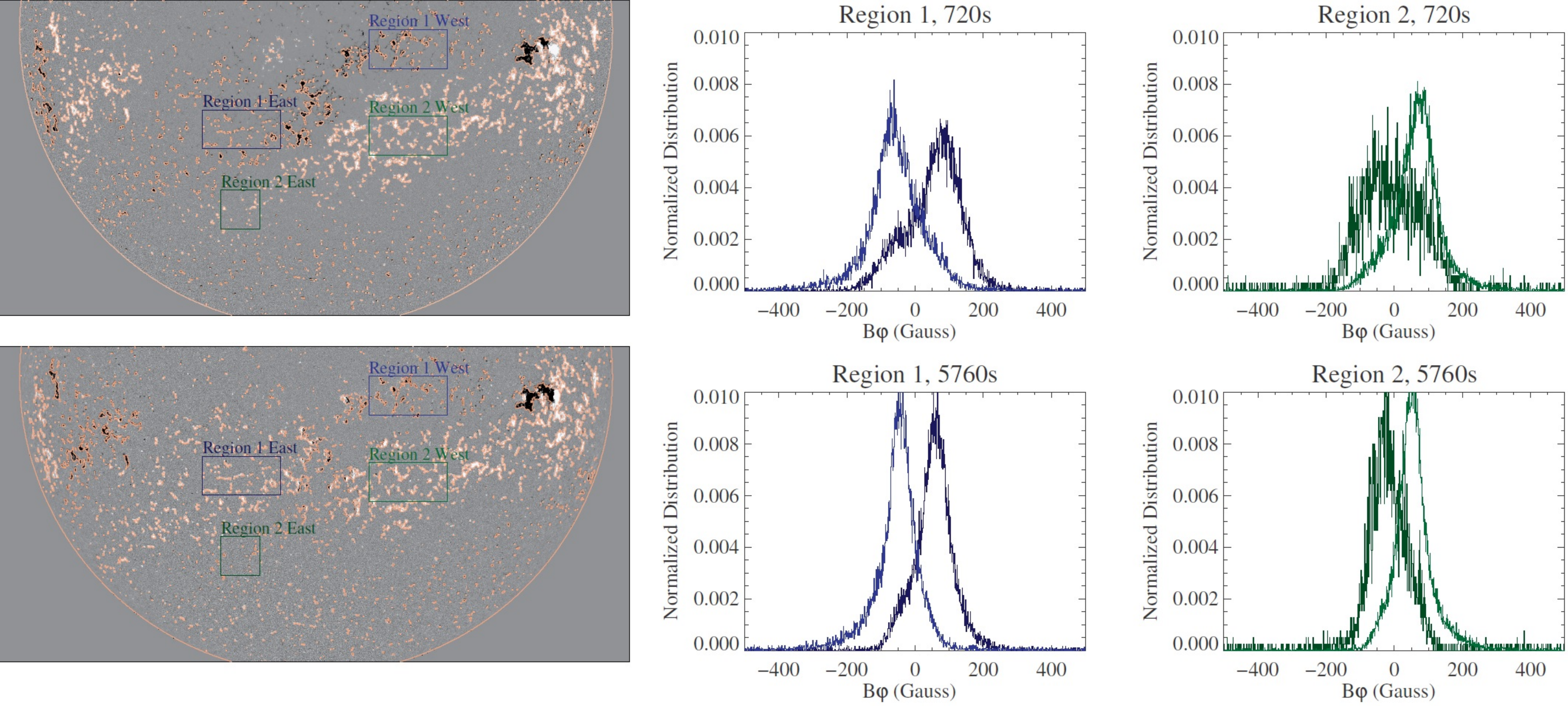}
\caption{
(Upper-left): half-disk image of radial B$_r$ field (white/black 
correspond to positive/negative polarity), scaled to $\pm$200 G. 
Outlines show pixels with the minimum-energy disambiguation 
applied (confid$\_$disambig keyword $\geq$ 60), which are included in the distributions shown in right columns.  
(Lower-left: half-disk image of zonal B$_\varphi$, scaled to 
$\pm$100 G. Region 1 (blue boxes) corresponds to area of negative 
polarity flux in decaying flux region. Region 2 (green boxes) 
corresponds to a similar area but of positive polarity flux. 
Boxes outlined by lighter/darker color are located West/East of the central meridian. (Middle column): distribution of 
the zonal (B$_\varphi$) component of the magnetic field
in the Region 1 for two (720~s and 5760~s) averaging.
(Right column): similar to middle column but for Region 2.
Note: salmon (light pinkish-orange) contours that outline pixels with the minimum-energy disambiguation mask underlying black/white magnetic fields. To see both the magnetic field and the contours, the PDF figure needs to be magnified by 300\% or more.
\label{fig:hmi}}
\end{figure}

Let's now consider whether lowering the noise level will mitigate this systematic 
effect. For this test, we selected HMI observations taken on 10 February 2015. That day
had a good representation of various solar features (plage, sunspots) situated at 
different distances from solar disk center. For the test, we employed filtergrams
that were processed using a much longer integration than normal (pipeline) 
magnetograms (5760 s vs. 720 s, courtesy Dr. X. Sun).
Integrated (average) filtergrams then were inverted 
using standard VFISV code, and disambiguated using the minimum 
energy disambiguation method with default pipeline settings. The averaged magnetogram is centered at 19:12:00 UT, and is 
shown in Figure \ref{fig:hmi} (left). Despite significant time averaging, we do not see
any obvious smearing of solar features. The S/N is improved when compared to the standard 
720s magnetograms, especially in areas with a weak polarization signal. 

To test the effect of lower noise on the inferred direction of the magnetic field vector in large-scale weak-signal areas, we select two decaying plage regions that extend across the central meridian, and sample the B$\varphi$ on either side (light/dark blue and green boxes in Figure \ref{fig:hmi}). We assume that the field vectors in these structures should not vary across the central meridian per se.
Figure \ref{fig:hmi} 
(right) shows the distribution of the zonal (East-West) magnetic field in the area of negative (blue colors) and positive (green colors) 
polarity flux. The distributions show a clear offset: for pixels situated East of 
central meridian, the mean is positive, while for pixels in the Western hemisphere, it is 
negative.
Such an offset supports the notion that a magnetic field situated in a region of decaying 
magnetic flux in the Eastern hemisphere, on average, shows inclination patterns away from the 
solar central meridian. Using observations with a lower noise level (5760~s) makes the 
distributions narrower, but the offset is still present. This behavior is in agreement 
with the explanation presented in Figure \ref{fig:demo}.

\section{Effects of azimuth disambiguation and filling factor} \label{sec:alpha}
For a simplified case of B$_\theta$ = 0, shown in Figure \ref{fig:demo}a, the transformation from longitudinal and transverse components, measured in the image plane at 
a heliographic longitude (central meridian distance) $\phi$, to heliographic components B$_r$ and B$_\varphi$ can be written as

\begin{equation}
B_r = B_{||}\cos\phi \pm B_\perp \sin\phi = 
\frac{C_{||}}{\alpha}~{\rm V}\cos\phi \pm
\frac{C_\perp}{\sqrt\alpha} \sqrt[4]{{\rm Q}^2 + {\rm U}^2} \sin\phi
\label{eq3}
\end{equation}

\begin{equation}
B_\varphi = - B_{||}\sin\phi \pm B_\perp \cos\phi = 
- \frac{C_{||}}{\alpha}~{\rm V}\sin\phi \pm
\frac{C_\perp}{\sqrt\alpha} \sqrt[4]{{\rm Q}^2 + {\rm U}^2} \cos\phi.
\label{eq4}
\end{equation}

For this simple configuration, the $\pm$ ambiguity in Equations \ref{eq3} and \ref{eq4} is resolved by requiring the
two components to have the same sign in Equation \ref{eq3} and consequently opposite
signs in Equation \ref{eq4}. The first component (Equation \ref{eq4})
is positive East of the central 
meridian ($\phi < 0$), and it is negative West of the central meridian. The second component is always 
positive. For the case when the field is mostly 
radial 
\begin{equation}
|\frac{-C_{||}~V}{\alpha}| >> \frac{C_\perp\sqrt[4]{Q^2+U^2}}{\sqrt{\alpha}}.
\label{eq:5}
\end{equation}
Under this condition, the sum of the two components, 
which represents the sign of B$_\varphi$, will be 
positive East of the central meridian and negative to 
the West. However, in the case of a mostly horizontal
field
\begin{equation}
|\frac{-C_{||}~V}{\alpha}| << \frac{C_\perp\sqrt[4]{Q^2+U^2}}{\sqrt{\alpha}}.
\label{eq:6}
\end{equation}
The sign of B$_\varphi$ will be always negative, 
independent of pixel location relative to the 
central meridian. This is in qualitative 
agreement with the observations mentioned in Section \ref{sec:intro} that the disagreement in the sign of the zonal (B$_\theta$) and/or meridional (B$\varphi$) components of HMI and VSM observations
was only observed in pixels with weaker linear 
polarization signals. In pixels with stronger 
linear polarization, the two instruments were in agreement with each other with respect to the sign of B$_\varphi$ and B$_\theta$ components.
Equations \ref{eq:5}-\ref{eq:6} suggest that the effect of sign 
reversal in B$_\varphi$ could also depend on magnetic filling factor 
$\alpha$. For some amplitudes of $\vert{-C_{||}~V}\vert$ and $C_\perp\sqrt[4]{Q^2+U^2}$ the inequity in Equation \ref{eq:5}
could change its sign if $\alpha$ is changed from being less then one to unity. The appendix provides an example of a test done with SOLIS/VSM data, which demonstrates that for $\alpha < 1$, B$_\varphi$ in three 
flux areas reverse their sign when the area crosses the central meridian. When $\alpha$ is set to unity, the same areas do not exhibit a sign reversal.

\section{Discussion} \label{sec:discuss}
We provide simple arguments and some observational evidence that a dichotomy in properties of
Zeeman polarization arising from longitudinal and transverse field components, and our ability to interpret
these signals especially in unresolved structures (non-unity filling factor), combined with the azimuthal disambiguation
may lead to erroneous 
conclusions about the orientation of the vector magnetic field, particularly in areas where the polarization signals are weak. 
The systematic patterns may depend on 
the amplitude of noise and whether the magnetic filling factor is resolved or assumed to be unity, and thus, observations from different instruments could result 
in slightly different orientation patterns (e.g., inclination of vector fields 
towards the solar poles or towards the solar equator). Pixels with stronger polarization 
signals are less affected (e.g., in sunspots, where $\alpha$ is relatively large, and where Stokes {\rm Q}, {\rm U}, and {\rm V} are typically strong), and thus, we expect that only pixels with weaker polarization signals will show the 
effect described above. Indeed, comparison of observations from SOLIS/VSM and SDO/HMI 
show a good agreement between the two instruments in areas of strong fields (e.g., 
sunspots), while in some areas of weak fields (e.g., plage), we see the zonal (East-West) 
and/or meridional (North/South) components having opposite sign. The amplitude of the 
effect will depend on the specific orientation of the magnetic field vector relative to 
line-of-sight, and thus the impact of this effect could vary across the solar disk in a somewhat complex way, making it 
difficult (or perhaps impossible) to correct, since the signal inherently originates from an unknown magnetic configuration. For an underlying radial-field orientation, the 
effect could be mitigated by equalizing the amplitude of the noise between the inferred longitudinal
and transverse field components. However, given the significant differences in noise levels 
between inferred B$_{||}$ and B$_\perp$ such noise equalization could be done only at the expense of 
increasing noise in the longitudinal field measurements (e.g., by using a modulation schema allowing much longer integration time for states corresponding to Stokes {\rm Q} and {\rm U} measurements). A test conducted using existing HMI data 
shows that simply improving S/N for both transverse and longitudinal fields does not completely eliminated the problem; 
while the amplitude of a systematic inclination decreases slightly, the overall effect still 
remains. 
We note that the systematic bias in the horizontal component of the magnetic field may, in principle, lead to 
the effects similar in appearance to ones that we discuss here.
While our simplified example discussed in Section \ref{sec:alpha} 
uses filling factor, it might be extremely difficult or even counterproductive
to model the exact magnetic and non-magnetic contribution in pixels outside sunspots.
The methods used to estimate the filling factor may inadvertently introduce 
additional errors related to the unknown difference in line profiles between 
magnetized and non-magnetized atmospheres in each pixel. The effects of the contribution function in spatial direction \citep[e.g., whether non-magnetized component is predominantly located on one side of a pixel or it is uniformly distributed around it as in][]{Harker2017} are also unknown. 
Perhaps, some of these issues may be addressed via spatially coupled inversion of spectro-polarimetric image data as in \citet{vanNoort2012}.
That technique, however, requires the knowledge of the Point-Spread-Function (PSF), which could be achieved either by measuring the atmospheric seeing during the observations \citep{vanNoort2017} or using adaptive optics. Thus, the method is not applicable for the existing SOLIS/VSM observations.

%
\begin{acknowledgements}

The authors thank Robert Cameron for his thoughtful 
comments, which helped improving this article.
We acknowledge help by Alexander Pevtsov in preparation of the manuscript.
This work is the result of discussions held at three 
working meetings on the ``Use of Vector Synoptic Maps for Modeling''. Funding for these meetings was 
provided by the hosting groups at the University of Oulu, Finland; the National Solar Observatory, USA; the Max Planck Institute for Solar System Research, Germany; 
and by NASA's Solar Dynamics Observatory (Dr. Dean Pesnell). 
We also acknowledge the financial support by the Academy of 
Finland to the ReSoLVE Centre of Excellence (project no. 307411).
AAP and LB acknowledge partial support by NASA grants 80NSSC17K0686 and 
NNX15AN43G. KDL acknowledges partial support from NASA/GSFC grant 
80NSSC19K0317.
This work utilizes SOLIS data obtained by the NSO Integrated Synoptic Program (NISP), managed 
by the National Solar Observatory. HMI data used here are courtesy of NASA/SDO and the HMI science teams.

\end{acknowledgements}

\begin{appendix}
\section{The effect of magnetic fill factor on the horizontal components of magnetic field}

\begin{figure}
\includegraphics[width=1.\columnwidth,clip=]{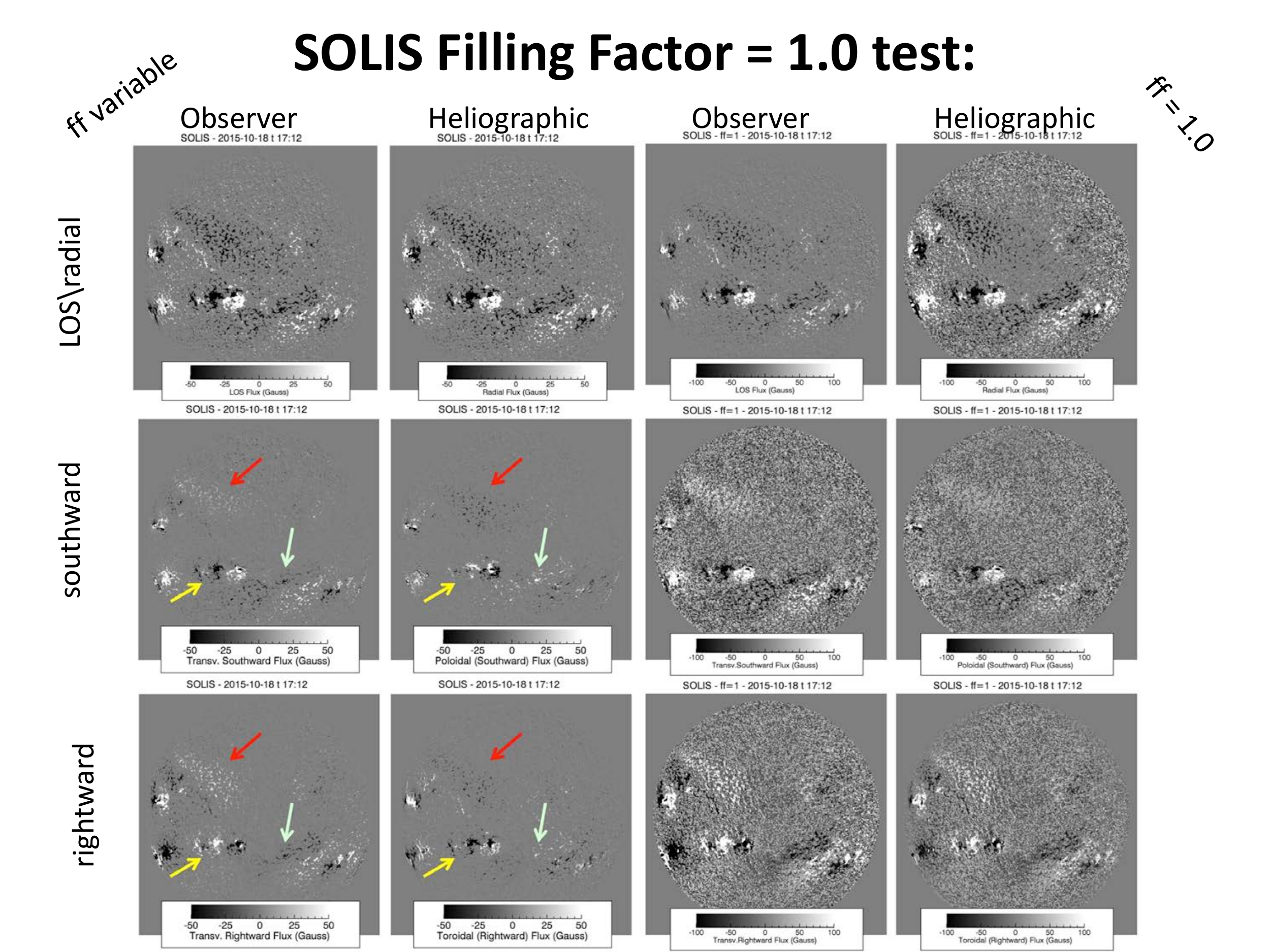}
\caption{Full disk images of B$_r$ (upper row), B$_\theta$ (middle) 
and B$_\varphi$ (low row). First and third columns show the vector 
field components in image coordinate system, while second and fourth 
columns are in the heliographic coordinate system. Three examples 
are marked by arrows of different color. Setting $\alpha$ to unity 
emphasizes pixels, which otherwise are below the noise threshold.
\label{fig:ff}}
\end{figure}

Figure \ref{fig:ff} provides an example of a test done with SOLIS/VSM data, which demonstrates that for $\alpha < 1$, B$_\varphi$ in three 
flux areas reverse its sign when the area crosses central meridian. 
When $\alpha$ is set to unity, the same areas do not exhibit the sign reversal.

\end{appendix}

\end{document}